\documentclass[doublecol]{epl2} 
\usepackage{amsfonts,amssymb,amsmath}
\usepackage{bm}
\usepackage{bbm}

\title{Properties of branching exponential flights in bounded domains}
\shorttitle{Properties of branching exponential flights in bounded domains} 

\author{A. Zoia\inst{1}, E. Dumonteil\inst{1} and A. Mazzolo\inst{1}}
\shortauthor{A. Zoia \etal}

\institute{
\inst{1} CEA/Saclay, DEN/DANS/DM2S/SERMA/LTSD, Gif-sur-Yvette, France}


\abstract{Branching random flights are key to describing the evolution of many physical and biological systems, ranging from neutron multiplication to gene mutations. When their paths evolve in bounded regions, we establish a relation between the properties of trajectories starting on the boundary and those starting inside the domain. Within this context, we show that the total length travelled by the walker and the number of performed collisions in bounded volumes can be assessed by resorting to the Feynman-Kac formalism. Other physical observables related to the branching trajectories, such as the survival and escape probability, are derived as well.}

\begin{document}

\maketitle

\section{Introduction}

\begin{figure}
\onefigure[scale=0.31]{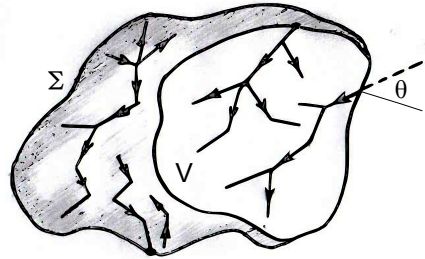}
\caption{A schematic representation of branching random walks in a bounded domain, starting on the surface $\Sigma$ or inside the volume $V$. For trajectories starting on the surface, the angle $\theta$ between the normal to the surface and the incoming trajectory obeys a density function depending on dimensionality.}
\label{fig1}
\end{figure}

Consider a single walker initially emitted from a point source. Once emitted, the walker undergoes a sequence of displacements, separated by collisions with the surrounding medium. When the scattering centers encountered by the travelling particle are spatially uniform, the inter-collision lengths are exponentially distributed~\cite{case}, with mean free path $\lambda$. The quantity $\sigma=1/\lambda$, proportional to the probability of particle-medium interaction along a straight line, is the total cross section: its value typically depends on the particle position and speed~\cite{case}. At each collision, the incident particle disappears, and $k$ particles (the descendants) are emitted with probability $p_k$, whose velocities are randomly redistributed in angle and intensity according to a given probability density, which in principle can vary as a function of the number $k$ of descendants~\cite{case}. Each descendant will then behave as the mother particle, and undergo a new sequence of displacements and collisions, giving thus rise to a branched structure, as illustrated in Fig.~\ref{fig1}. Branching random walks as described above lie at the heart of physical and biological modeling, and are key to the description of neutron transport in multiplying media and nucleon cascades, spread of epidemics, diffusion of reproducing bacteria, and mutation-propagation of genes, just to name a few~\cite{harris, pazsit, jagers}.

A central question for random walks is to determine the occupation statistics of the stochastic paths in a given region $V$ of the phase space: for branching or non-branching Brownian motion, for instance, this is intimately related to such issues as residence times and first-passage properties~\cite{condamin_benichou, condamin, grebenkov, berezhkovskii_zaloj, agmon_lett, sawyer, cox}. For exponential flights, the occupation statistics is naturally defined in terms of two observables: the number of occurred collisions in the volume $V$ and the total length travelled in $V$~\cite{case}. In principle, one would be interested in assessing the full distribution of these two quantities: this unfortunately turns out to be a formidable task, even for very simple geometries, and analytical results are seldom available~\cite{zdm_pre, zdm_prl, zdm_epl, ZDMS_arxiv}.

A somewhat simpler approach consists in deriving formulas for the moments of the distributions, which are sometimes amenable to exact results. For  exponential flights, important findings have been reported in recent years concerning trajectories starting either on the boundary or inside a given bounded volume, in the special case $p_1=1$, i.e., the well-known Pearson random walk~\cite{mazzolo_jphysA, blanco_europhys, mazzolo_europhys, benichou_europhys, dubi}. In particular, it has been shown that the average travelled length for walkers starting from the boundary depends only on the ratio of the volume over the surface, which is a rather counter-intuitive result. In this Letter we extend these findings in two directions, based on the Feynman-Kac approach~\cite{kac}: first, we include more generally absorption and branching mechanisms; second, we show that other physical observables associated to the process, such as escape and survival probabilities, can be easily derived within the same formalism.

\section{Travelled lengths}

Consider a bounded domain of non-zero volume $V$ (with finite diameter) and non-zero measurable surface $\Sigma=\partial V$. An exponential flight started in $V$ or on its boundary will end when all descendants have been either absorbed in $V$ or escaped from $\Sigma$. In order to keep notation simple, yet retaining the key features of the process, we will assume in the following that particles evolve at constant speed $v$, descendants are emitted isotropically and independently from each other, and $\sigma$ is constant over the volume $V$. To start with, we denote by ${L}^m ({\mathbf r}_0,\boldsymbol{\omega}_0,t )$ the $m$-th moment of the total length travelled in $V$ by the particle and all its descendants when observed up to time $t$, starting from a single walker in ${\mathbf r}_0$ with direction $\boldsymbol{\omega}_0 $ at time $t=0$. It is possible to characterize $L^m$ by resorting to the Feynman-Kac formalism~\cite{ZDMS_arxiv}, which yields the recursive formula
\begin{eqnarray} 
	 \frac{1}{v}\frac{\partial}{\partial t}{L}^m ({\mathbf r}_0,\boldsymbol{\omega}_0,t )
	= \sigma \sum_{j=2}^m \nu_j {\cal B}_{m,j} \left[ \langle L^i \rangle_{\Omega}(\mathbf r_0,t) \right] \nonumber \\
	+ {\cal L}^*{L}^m ({\mathbf r}_0,\boldsymbol{\omega}_0,t )
	+ m \mathbbm{1}_{V} {L}^{m-1} ({\mathbf r}_0,\boldsymbol{\omega}_0,t ),
\label{eq.moment_kac_ell_L}
\end{eqnarray}
for $m \ge 1$, starting with ${L}^m ({\mathbf r}_0,\boldsymbol{\omega}_0,0 )=0 $ and ${L}^0 ({\mathbf r}_0,\boldsymbol{\omega}_0,t )=1 $ from normalization. Here
\begin{equation}
	{\cal L}^* = \boldsymbol{\omega}_0 \cdot \nabla_{{\mathbf r}_0} - \sigma + \sigma \nu_1 \int \frac{d\boldsymbol{\omega}_0}{\Omega_d}
\end{equation}
is the backward transport operator~\cite{case}, $\nu_1=\langle k\rangle = \sum_k k p_k$ being the average number of secondary particles per collision, and $\Omega_d=2\pi^{d/2}/\Gamma(d/2)$ the surface of the unit sphere in dimension $d$. Furthermore, we denote by
\begin{equation}
 \langle h \rangle_{\Omega}(\mathbf r_0,t) \equiv \int \frac{d\boldsymbol{\omega}_0}{\Omega_d} h ({\mathbf r}_0,\boldsymbol{\omega}_0,t).
\end{equation}
the average over directions, and by $\mathbbm{1}_{V}$ the marker function of the volume $V$, i.e., $\mathbbm{1}_{V}=1$ when ${\mathbf r}_0$ belongs to $V$, and $\mathbbm{1}_{V}=0$ elsewhere. Finally, ${\cal B}_{m,j} \left[z_{i}\right]={\cal B}_{m,j} \left[z_{1},z_{2},\cdots,z_{m-j+1}\right]$ stand for the Bell's polynomials~\cite{pitman_book}, $\nu_j=\langle k(k-1)...(k-j+1)\rangle$ being the falling factorial moments of the descendant number, starting with $\nu_0=1$. When trajectories are observed up to a time $t$ much larger than the characteristic time scale of the system dynamics, we can define the stationary moments ${L}^m ({\mathbf r}_0,\boldsymbol{\omega}_0) = \lim\limits_{t \to +\infty} {L}^m ({\mathbf r}_0,\boldsymbol{\omega}_0,t ) $, provided that the limit exists. Intuitively, this condition is satisfied when the particle losses due to absorptions and leakages from the boundaries are larger than the gain due to population growth, which is always the case if $\nu_1 \le 1$~\cite{harris}. When $\nu_1 > 1$, this typically amounts to further requiring that $V$ is below some critical size $V_c$~\cite{case}. In the following we will assume that $V<V_c$, unless differently specified: the time derivative in Eq.~\ref{eq.moment_kac_ell_L} then vanishes at large times, and we get
\begin{eqnarray}
	{\cal L}^*{L}^m ({\mathbf r}_0,\boldsymbol{\omega}_0) + m \mathbbm{1}_{V} {L}^{m-1} ({\mathbf r}_0,\boldsymbol{\omega}_0) \nonumber \\
	+\sigma \sum_{j=2}^m \nu_j {\cal B}_{m,j} \left[ \langle L^i \rangle_{\Omega}(\mathbf r_0) \right]=0,
\label{eq.moment_kac_ell_L_stat}
\end{eqnarray}
with the boundary conditions ${L}^m ({\mathbf r}_0 \in \Sigma,\boldsymbol{\omega}_0)=0$ when $\boldsymbol{\omega}_0$ is directed outward.

Following~\cite{benichou_europhys}, our aim is now to average Eq.~\ref{eq.moment_kac_ell_L_stat} over the starting position and direction of the walker. As the domain $V$ is bounded, we can safely define the probability measures for trajectories born in the domain and for those starting on the surface. Choosing the starting coordinates uniformly distributed inside $V$ imposes the uniform volume probability measure
\begin{equation}
\frac{d\boldsymbol{\Omega}}{\Omega_d}\frac{dV}{V},
\end{equation}
where $d\boldsymbol{\Omega}$ is the solid angle element~\footnote{A $\mu-$ or equivalently $\nu-$randomness in the language of stochastic geometry~\cite{santalo, mazzolo_jphysA}.}. Similarly, an isotropic incident flux uniformly distributed on the frontier $\Sigma$ imposes the surface probability measure
\begin{equation}
\eta_d \frac{d\boldsymbol{\Omega}}{\Omega_d}\frac{d\Sigma}{\Sigma} (\boldsymbol{\Omega} \cdot \bf n),
\end{equation}
where $\eta_d = \sqrt{\pi}(d-1)\Gamma((d-1)/2)/\Gamma(d/2)$ is a dimension-dependent normalization constant, equal to twice the inverse of the average height of the $d$-dimensional unit shell, and $\boldsymbol{n}$ is the normal entering the surface~\footnote{A $\mu-$randomness~\cite{santalo, mazzolo_jphysA}. The term $\cos\theta=(\boldsymbol{\Omega} \cdot \mathbf{n})$ implies that in polar coordinates trajectories starting on the surface must enter the domain with density $\theta = \arcsin(2 \xi -1)$ in two dimensions and $\theta = 1/2 \arccos(1- 2 \xi)$ in three dimensions, $\xi$ being uniformly distributed in $(0,1]$; see Fig.~\ref{fig1}.}. By means of such probability measures, for any function $h({\mathbf r}_0,\boldsymbol{\omega}_0)$ we define its volume average $\langle h \rangle_V$ and its surface average $\langle h \rangle_\Sigma$, respectively, by
\begin{eqnarray}
\langle h \rangle_V &=& \int \frac{d \mathbf{r}_0 }{V} \int \frac{d\boldsymbol{\omega}_0}{\Omega_d} h(\mathbf{r}_0,\boldsymbol{\omega}_0),\nonumber \\
\langle h \rangle_\Sigma &=& \eta_d \int \frac{d\Sigma }{\Sigma} \int \frac{d\boldsymbol{\omega}_0}{\Omega_d} (\boldsymbol{\omega}_0 \cdot \mathbf{n}) h(\mathbf{r}_0,\boldsymbol{\omega}_0).
\label{eq.mean_surf}
\end{eqnarray}
We integrate then Eq.~\ref{eq.moment_kac_ell_L_stat} uniformly over all possible initial positions and directions (taking into account the isotropy property), and apply the Gauss divergence theorem. This yields the recursive formula
\begin{eqnarray} 
\langle L^m \rangle_\Sigma & = & \eta_d \frac{V}{\Sigma} \Big[ m \langle L^{m-1} \rangle_V + \sigma \left(\nu_1 - 1\right) \langle L^m \rangle_V  \nonumber \\
	& + & \sigma \sum_{j=2}^m \nu_j  \langle  {\cal B}_{m,j} \left[ \langle L^i \rangle_{\Omega}(\mathbf r_0) \right] \rangle_{V} \Big].
\label{eq.mean_vol_surf}
\end{eqnarray}
Equation~\ref{eq.mean_vol_surf} relates the $m$-th moment of trajectories starting on the surface to the different moments (up to order $m$) of trajectories born inside the volume, and as such extends to branching random flights the formulas
\begin{equation}
\langle L \rangle_\Sigma = \eta_d \frac{V}{\Sigma} \mathrm{~~and~~} \langle L^{m-1} \rangle_V = \frac{\langle L^m \rangle_\Sigma}{m \langle L\rangle_\Sigma} \mathrm{~~for~~} m \ge 1
\label{eq.diffusive}
\end{equation}
previously obtained for Pearson random walks~\cite{blanco_europhys, mazzolo_europhys, benichou_europhys}. In particular, from Eq.~\ref{eq.mean_vol_surf} the average length ($m=1$) reads
\begin{equation}
 \langle L \rangle_\Sigma = \eta_d \frac{V}{\Sigma} \left[ 1 + \sigma \left(\nu_1 - 1\right) \langle L \rangle_V \right],
\label{eq.mean}
\end{equation}
which generalizes the celebrated Cauchy's formula $\langle L \rangle_\Sigma = \eta_d \frac{V}{\Sigma}$ (also known as the mean chord length property), originally established for random straight lines drawn from the surface of the volume~\cite{santalo} and recently shown to rather surprisingly apply also to Pearson random flights~\cite{dubi, blanco_europhys, mazzolo_europhys, benichou_europhys}.

The term $O = \langle L \rangle_\Sigma / \lambda$ is a measure of the opacity of the volume, in that it expresses the ratio between the average length travelled in $V$ when an isotropic particle flux is imposed at the surface $\Sigma$ and the mean free path. Another quantity of interest is $\chi = \langle L \rangle_\Sigma / (\eta_d V/\Sigma)$, which is the ratio between the average length travelled in the actual medium $V$ and the length that the particle would elapse if $V$ were empty and the paths were straight lines (the meaning of the denominator stems from Cauchy's formula).

In general, Eq.~\ref{eq.mean} depends on the fine details of the process and of the geometry, since the term $\langle L \rangle_V$ is not universal. However, when the underlying branching process has $\nu_1=1$, Eq.~\ref{eq.mean} yields precisely the Cauchy's formula. In this case, thus, the quantity $\langle L \rangle_\Sigma$ would depend only on the geometrical ratio $V/\Sigma$ and not on the specific details of the random walk. In particular, $\langle L \rangle_\Sigma$ would be independent of the characteristic jump size $\lambda$. This simple property unfortunately does not carry over to higher moments of the travelled length. Indeed, for $m=2$ we have ${\cal B}_{2,2} [z_1,z_2]=z^2_1$, and Eq.~\ref{eq.mean_vol_surf} then gives
\begin{equation}
\langle L^2 \rangle_\Sigma \!=\! \eta_d \frac{V}{\Sigma} \left[ 2 \langle L \rangle_V \!+\! \sigma \left(\nu_1 \!-\! 1\right) \langle L^2 \rangle_V  + \sigma \nu_2 \langle  \langle L \rangle_{\Omega}^2(\mathbf r_0) \rangle_{V} \right]\! \nonumber.
\end{equation}
Bell's polynomials in Eq.~\ref{eq.mean_vol_surf} are the signature of branching, and for $m \ge 2$ introduce some extra non-vanishing terms ($\nu_2 > 0$) with respect to Eq.~\ref{eq.diffusive} even when $\nu_1=1$.

\begin{figure}
\onefigure[scale=0.25]{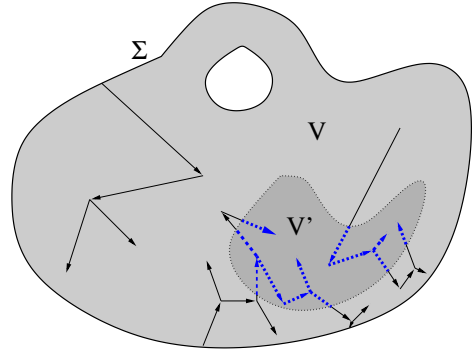}
\caption{Branching random flights born in the volume $V$ (or on the surface $\Sigma$) and traversing a sub-domain $V'$ of $V$. The portion of the travelled trajectories spent in $V'$ is displayed as blue dashed lines.}
\label{fig2}
\end{figure}

In the absence of branching, i.e., when random flights can be either scattered or absorbed, with $p_0+p_1=1$ and $p_k=0$ for $k \ge 2$, explicit relations for the probability density functions of the travelled length can be also derived. Under these hypotheses, Eq.~\ref{eq.mean_vol_surf} reduces to
\begin{eqnarray} 
\langle L^m \rangle_\Sigma = \eta_d \frac{V}{\Sigma} \left[ m \langle L^{m-1} \rangle_V - \sigma_0 \langle L^m \rangle_V \right],
\label{eq.mean_vol_surf_nobranching}
\end{eqnarray}
where $\sigma_0 = p_0 \sigma$ corresponds to the absorption cross section. In presence of absorption, then, $\langle L^m \rangle_\Sigma < \eta_d V/\Sigma$, as expected. We denote by $f(l)$ and $g(r)$ the probability density of the total travelled length for a trajectory started on the surface or inside the volume, respectively. Then, Eq.~\ref{eq.mean_vol_surf_nobranching} can be identically rewritten as
\begin{equation}
\int_0^{\infty} l^m f(l) dl = \eta_d \frac{V}{\Sigma} \int_0^{\infty}\left[ \frac{m}{r}-\sigma_0\right]r^{m} g(r) dr.
\label{eq_integral}
\end{equation}
Integrating Eq.~\ref{eq_integral} by parts and using normalization $\int_0^{\infty} g(r)dr = 1$ yields the relation
\begin{equation}
 g(r) = \frac{1}{\eta_d \frac{V}{\Sigma}} \left[1 +  \eta_d \frac{V}{\Sigma}\sigma_0 - \int_0^r f(l) e^{\sigma_0 l} dl \right] e^{-\sigma_0 r},
\label{eq.differential_solution}
\end{equation}
between the two densities $f(l)$ and $g(r)$. In the limit of purely diffusive processes ($p_0 \to 0$), Eq.~\ref{eq.differential_solution} reduces to
\begin{equation}
 g(r) = \frac{1}{\eta_d \frac{V}{\Sigma}} \int_r^{\infty} f(l)dl,
\end{equation}
a relation originally established for straight paths~\cite{dixmier} and later extended to Pearson walks~\cite{mazzolo_europhys}.

\section{Excursions in sub-domains}

Considerable efforts have been devoted to the study of the occupation statistics of some sub-domain $V'$ included in $V$. This issue has been thoroughly investigated, e.g., in the context of residence times for Brownian motion (with or without branching; see, e.g.,~\cite{grebenkov, berezhkovskii_zaloj, agmon_lett, sawyer, cox}) and Pearson walks~\cite{benichou_europhys}. Consider a branching exponential flight emitted in $V$: the particle and its descendants may enter $V'$, spend some time inside, branch, possibly die in $V'$ or escape, then re-enter $V'$, and so on, as illustrated in Fig.~\ref{fig2}. The total length travelled in $V'$ can be straightforwardly assessed by resorting to the Feynman-Kac formalism mentioned above. Indeed, its moments ${L'^m}({\mathbf r}_0,\boldsymbol{\omega}_0)$ satisfy Eq.~\ref{eq.moment_kac_ell_L_stat}, the marker function being restricted to the sub-domain $V'$. Then, averaging over all angles and positions inside $V$ yields
\begin{eqnarray}
\langle L'^m \rangle_\Sigma & = & \eta_d \frac{V}{\Sigma} \Big[ m \frac{V'}{V}\langle L'^{m-1} \rangle_{V'} + \sigma \left(\nu_1 - 1\right) \langle L'^m \rangle_V  \nonumber \\
	& + & \sigma \sum_{j=2}^m \nu_j  \langle  {\cal B}_{m,j} \left[ \langle L'^i \rangle_{\Omega}(\mathbf r_0) \right] \rangle_{V} \Big].
\end{eqnarray}
For $m=1$ we have in particular
\begin{equation}
  \langle L' \rangle_\Sigma = \eta_d \frac{V}{\Sigma} \left[ \frac{V'}{V} + \sigma \left(\nu_1 - 1\right) \langle L' \rangle_V \right].
\label{eq.mean.sub}
\end{equation} 
As a consequence, for trajectories starting on the surface, the ratio between the length travelled in $V'$ and that travelled in $V$ reads
\begin{equation}
\frac{\langle L' \rangle_\Sigma}{\langle L \rangle_\Sigma}  = \frac{V'}{V} \left[ \frac{ 1 + \frac{V}{V'} \sigma \left(\nu_1 - 1\right) \langle L' \rangle_V}{ 1 + \sigma \left(\nu_1 - 1\right) \langle L \rangle_V} \right].
\label{eq.mean.fraction}
\end{equation}
This ratio generally depends on the geometry as well as on the walk features. However, for branching flights with $\nu_1=1$, we have $\langle L' \rangle_\Sigma/\langle L \rangle_\Sigma=V'/V$, i.e., we recover the elegant ergodic-type property that applies to Pearson walks~\cite{benichou_europhys}.

\section{Number of collision events}

The stationary $m$-th moment $N^{m}$ of the number of collisions in $V$ performed by a branching exponential flight starting from ${\mathbf r}_0$ in direction $\boldsymbol{\omega}_0 $ can be also assessed by resorting to the Feynman-Kac formalism: as shown in~\cite{ZDMS_arxiv}, $N^{m}$ satisfies
\begin{eqnarray}
{\cal L}^* N^{m}  ({\mathbf r}_0,\boldsymbol{\omega}_0) + \sigma \sum_{j=2}^m \nu_j {\cal B}_{m,j} \left[ \langle N^{i} \rangle_\Omega \right]+\nonumber \\
\sigma \mathbbm{1}_{V}\sum_{k=1}^m  {m \choose k}  \sum_{j=0}^{m-k} \nu_j {\cal B}_{m-k,j} \left[ \langle N^{i} \rangle_\Omega \right]=0,
\label{moment_kac_n_stat}
\end{eqnarray}
which closely resembles Eq.~\ref{eq.moment_kac_ell_L_stat}. At the boundaries, we have ${N}^m ({\mathbf r}_0 \in \Sigma,\boldsymbol{\omega}_0)=0$ when $\boldsymbol{\omega}_0$ is directed outward. Then, by averaging Eq.~\ref{moment_kac_n_stat} over starting positions and directions in $V$ we get
\begin{eqnarray} 
\langle N^m \rangle_\Sigma = \eta_d \frac{V}{\Sigma} \sigma \Big[ \sum_{k=1}^m  {m \choose k}  \sum_{j=0}^{m-k} \nu_j \langle {\cal B}_{m-k,j} \left[ \langle N^i \rangle_{\Omega}(\mathbf r_0) \right] \rangle_{V} \nonumber \\
+\left(\nu_1 - 1\right) \langle N^m \rangle_V  + \sum_{j=2}^m \nu_j  \langle  {\cal B}_{m,j} \left[ \langle N^i \rangle_{\Omega}(\mathbf r_0) \right] \rangle_{V}
\Big] \nonumber,
\label{eq.mean_vol_surf_N}
\end{eqnarray}
which generalizes the results previously found for Pearson walks~\cite{mazzolo_ane}. For the average collision number we have
\begin{equation}
\langle N \rangle_\Sigma = \eta_d \frac{V}{\Sigma} \sigma \left[ 1 + \left(\nu_1 - 1\right) \langle N \rangle_V \right].
\end{equation}
Now, from exponential flights being a Markovian process it follows that $\langle N \rangle_V = \sigma \langle L \rangle_V$~\cite{ZDMS_arxiv}. Hence, we have also  $O=\sigma \langle L \rangle_\Sigma = \langle N \rangle_\Sigma $, which amounts to saying that the opacity can be expressed in terms of the mean number of collisions in $V$. Similarly as done for the lengths, the number of collisions in a sub-domain $V'$ can again be computed by using $\mathbbm{1}_{V'} $ in Eq.~\ref{moment_kac_n_stat}.

For non-branching flights, $\nu_j=0$ for $j \geq 2$ and, from ${\cal B}_{j,1} \left[z_{i} \right]= z_j $ and ${\cal B}_{j,0} \left[z_{i} \right]= 0 $ for $j \ge 1$, we have
\begin{equation}
\langle N^m \rangle_\Sigma = \eta_d \frac{V}{\Sigma} \sigma \left[ p_1 \sum_{k=0}^{m} {m \choose k} \langle N^{m-k} \rangle_{V} - \langle N^m \rangle_V +p_0 \right].\nonumber
\end{equation}
By making use of the binomial formula, we finally get
\begin{equation}
\langle N^m \rangle_\Sigma = \eta_d \frac{V}{\Sigma} \sigma \left[ p_1 \langle (N+1)^{m} \rangle_{V} -\langle N^{m}\rangle_{V} +p_0\right].
\label{eq.mean_vol_surf_N_diffusion}
\end{equation}
In the absence of branching, it is also possible to explicitly derive the collision probabilities. We denote by $f(i)$ and $g(j)$ the collision number probability for a trajectory started on the surface or inside the volume, respectively. Then, Eq.~\ref{eq.mean_vol_surf_N_diffusion} can be identically rewritten as
\begin{equation}
\sum_{i=0}^{\infty} i^m f(i) = \eta_d \frac{V}{\Sigma} \sigma \left[ \sum_{j=0}^{\infty} \left[p_1(j+1)^m -j^m \right] g(j) +p_0\right].\nonumber
\end{equation}
By equating the terms of the series, and imposing that the relation holds for arbitrary $m\ge 1$, we get then
\begin{equation}
f(j) = \eta_d \frac{V}{\Sigma} \sigma \left[ p_1 g(j-1) - g(j) + p_0 \delta_{j,1} \right],
\end{equation}
$\delta_{i,j}$ being the Kronecker delta. Resumming over $j$ yields in particular $f(0) = 1 - \eta_d \frac{V}{\Sigma} \sigma g(0)$. For Pearson walks, $p_0 \to 0$ and we obtain
\begin{equation}
f(j) = \eta_d \frac{V}{\Sigma} \sigma \left[ g(j-1) - g(j) \right].
\end{equation}

\section{Escape probability}

When the volume $V$ is bounded, the probability $R({\mathbf r}_0,\boldsymbol{\omega}_0)$ that a trajectory starting from ${\mathbf r}_0$ in direction $\boldsymbol{\omega}_0 $ never visits the exterior of $V$ satisfies~\cite{ZDMS_arxiv}
\begin{equation}
-\boldsymbol{\omega}_0 \cdot \nabla_{{\mathbf r}_0}R({\mathbf r}_0,\boldsymbol{\omega}_0) + \sigma R({\mathbf r}_0,\boldsymbol{\omega}_0)= \sigma \mathbbm{1}_{V} G\left[ \langle R \rangle_\Omega \right],
\label{eq.R} 
\end{equation}
where $G[z]=\sum_k p_k z^k$ is the generating function associated to the descendant number distribution $p_k$ and $R ({\mathbf r}_0 \in \Sigma,\boldsymbol{\omega}_0)=0$ when $\boldsymbol{\omega}_0$ is directed outward. By definition, the quantity $Q=1-R$ represents the escape probability from the volume. Then, averaging Eq.~\ref{eq.R} over all initial positions and directions yields
\begin{equation} 
\langle R \rangle_\Sigma = \eta_d \frac{V}{\Sigma} \sigma \left[ \langle G\left[\langle R \rangle_\Omega  \right] \rangle_V - \langle R \rangle_V\right].
\label{eq.mean_R}
\end{equation}
The terms $\langle R \rangle_V $ and and $\langle R \rangle_\Sigma$ have a simple probabilistic meaning, namely, the probability that a particle born uniformly and isotropically in the volume $V$, or entering the body isotropically from the boundary, respectively, is absorbed with all its descendants in $V$~\cite{case}. Similarly, $\langle R \rangle_{\Omega} (\mathbf r_0) $ represents the probability that a particle born isotropically at ${\mathbf r}_0$ is absorbed (with all its descendants) in $V$. Equation~\ref{eq.mean_R} relates therefore the spatial behavior of the probability $R$ to the probabilities $p_k$, via $G[z]$.

When walkers can not be absorbed in the domain ($p_0=0$), trajectories must necessarily escape from the boundaries, and we have $\langle R \rangle_\Sigma = 0$. This rather intuitive result can be understood as follows: developing Eq.~\ref{eq.mean_R} and using the normalization $\sum_{k=0}^{\infty} p_k =1$ yields
\begin{equation} 
\langle R \rangle_\Sigma = \eta_d \frac{V}{\Sigma} \sigma \sum_{k=2}^{\infty} p_k \left[ \langle \langle R \rangle_\Omega^k \rangle_V - \langle R \rangle_V \right].
\label{eq.dev_R} 
\end{equation}
Since $\langle R \rangle_\Omega^k \leq \langle R \rangle_\Omega$ ($\langle R \rangle_\Omega$ is a probability), we immediately get that $\langle R \rangle_\Sigma \leq 0$: then, the probability that a particle entering the body isotropically from the boundary is absorbed with all its descendants in $V$ must vanish, as expected. The same applies to $\langle R \rangle_V$.

In the absence of branching, $G[z] = p_0 + p_1z$ with $p_0+p_1=1$, and we get
\begin{equation} 
	\langle R \rangle_\Sigma  =  \eta_d \frac{V}{\Sigma} \sigma_0 \left[ 1 - \langle R \rangle_V \right],
\end{equation}
a $d$-dimensional generalization of a theorem originally derived for purely absorbing media~\cite{case} and extended to diffusive and absorbing media in three dimensions in~\cite{mazzolo_europhys}.

\section{Survival probability}

A fundamental quantity for random walks is the survival probability $S_t({\mathbf r}_0,\boldsymbol{\omega}_0)$, namely, the probability that at time $t$ at least one particle is still in $V$. When $V < V_c$, for long times $S_t({\mathbf r}_0,\boldsymbol{\omega}_0) \to 0$, which physically means that the combined effects of absorptions in $V$ and leakages through $\Sigma$ are sufficient to compensate the population growth due to branching (if any), and trajectories almost surely go to extinction. However, when $\nu_1 >1$ and $V>V_c$, branching paths have a finite probability of surviving indefinitely in $V$, and there exists a non-trivial limit $S_t({\mathbf r}_0,\boldsymbol{\omega}_0) \to S({\mathbf r}_0,\boldsymbol{\omega}_0) > 0$, depending on the starting position and direction. For branching Brownian motion, for instance, finding the asymptotic survival probability is a long standing issue~\cite{derrida, derrida_barrier}. For branching exponential flights in bounded domains, when $V>V_c$ the probability of ultimate survival $S({\mathbf r}_0,\boldsymbol{\omega}_0)$ satisfies~\cite{ZDMS_arxiv}
\begin{equation}
-\boldsymbol{\omega}_0 \cdot \nabla_{{\mathbf r}_0}S({\mathbf r}_0,\boldsymbol{\omega}_0) + \sigma S({\mathbf r}_0,\boldsymbol{\omega}_0)= \sigma F[\langle S \rangle_\Omega],
\label{extinction_prob}
\end{equation}
where $F[z]=\sum_{k=1}^\infty \alpha_k z^k$, with $\alpha_k=(-1)^{k+1} \nu_k / k! $. At the boundaries, $S$ must vanish when $\boldsymbol{\omega}_0$ is directed towards the exterior of $V$. Averaging Eq.~\ref{extinction_prob} over all initial positions and directions yields then
\begin{equation} 
\langle S \rangle_\Sigma = \eta_d \frac{V}{\Sigma} \sigma \left[ \langle F\left[\langle S \rangle_\Omega  \right] \rangle_V  - \langle S \rangle_V \right].
\label{survival_probability}
\end{equation}
Unfortunately, the complex nature of the alternating series in $F[z]$ seems to prevent from drawing general conclusions based on Eq.~\ref{survival_probability}.

\section{Perspectives}

The approach proposed in this Letter based on the Feynman-Kac formalism allows the volume-averaged properties of branching exponential flights to be derived by relying upon a minimal number of simplifying hypotheses. The results presented here are fairly general, and as such apply to a broad class of physical and biological systems. Moreover, no assumptions are made concerning the shape of the domain, and our findings remain valid for non-convex bodies, including for instance domains with holes. Further investigations are ongoing in three directions: including the effects of anisotropies and heterogeneities, both in the source terms and at the collision events~\cite{zdm_pre_operator}, allowing for other (possibly unbounded) jump distributions, such as L\'evy Flights~\cite{zrk_lf}, and taking into account reflective or mixed boundary conditions~\cite{benichou_europhys}.

\acknowledgements The authors wish to thank S.~Mohamed for carefully checking formulas by Monte Carlo simulation.

\end{document}